\documentclass[12pt,a4paper]{article}
\usepackage{amsfonts}
\usepackage[onehalfspacing]{setspace}
\usepackage[top=1.25in, bottom=1.25in, left=1.25in, right=1.25in]{geometry}

\newenvironment{proof}[1][Proof]{\noindent\textbf{#1.} }{\ \rule{0.5em}{0.5em}}
\input{tcilatex}
\begin{document}

\title{Posterior Probabilities: Dominance and Optimism}
\author{Sergiu Hart\thanks{%
The Hebrew University of Jerusalem (Federmann Center for the Study of
Rationality, Department of Economics, and Institute of Mathematics).\quad 
\emph{E-mail}: \texttt{hart@huji.ac.il} \quad \emph{Web site}: \texttt{%
http://www.ma.huji.ac.il/hart}} \and Yosef Rinott\thanks{%
The Hebrew University of Jerusalem (Federmann Center for the Study of
Rationality, and Department of Statistics).\quad \emph{E-mail}: \texttt{%
yosef.rinott@mail.huji.ac.il} \quad \emph{Web site}: \texttt{%
http://pluto.huji.ac.il/\symbol{126}rinott}}}
\date{April 27, 2020}
\maketitle

\begin{center}
{\Large \ \textbf{Supplementary Material }}
\end{center}

\section{Dominance}

We provide some details on the example at the end of Section 1 of the paper.
The simple remark below is useful here (as well as in the next section).

\bigskip

\noindent \textbf{Remark.} In the binary case where $X$ and $Y$ take only
two values, say $v_{1}<v_{2},$ likelihood ratio dominance is easily seen to
be equivalent to the high value having a higher probability; i.e., $X\geq _{%
\mathrm{lr}}Y$ if and only if $\mathbb{P}(X=v_{2})\geq \mathbb{P}(Y=v_{2})$
(which is also equivalent to first-order stochastic dominance).

\bigskip

\noindent \textbf{Example.} Suppose that the set of states is $\Theta
=\{\alpha ,\beta ,\gamma \},$ the prior $\pi $ is uniform on $\Theta $
(i.e., each state has probability $1/3),$ and $\Gamma =\{\beta ,\gamma \}.$
Let $S=\{0,1\}$ be the set of signals, with signaling probabilities $\sigma
(s|\theta )$ as given in the table below (the last two columns then give the
total probability $\mathbb{P}(s)$ of each signal $s$ and the posterior $%
Q_{\Gamma }(s)\equiv \mathbb{P}(\Gamma |s))$:%
\[
{\renewcommand{\arraystretch}{2.8}\setlength{\tabcolsep}{7pt}}%
\begin{tabular}{c||ccc||c|c||c}
$s$ & $\sigma (s|\alpha )$ & $\mathbb{\sigma }(s|\beta )$ & $\mathbb{\sigma }%
(s|\gamma )$ & $\mathbb{P}(s)$ & $\mathbb{P}^{\Gamma }(s)$ & $Q_{\Gamma }(s)$
\\ \hline\hline
$1$ & $3/8$ & $3/4$ & $1/2$ & $13/24$ & $5/8$ & $10/13$ \\ 
$0$ & $5/8$ & $1/4$ & $1/2$ & $11/24$ & $3/8$ & $6/11$%
\end{tabular}%
\]%
Proposition 1 gives $(Q_{\Gamma },\mathbb{P}^{\Gamma })>_{\mathrm{lr}%
}(Q_{\Gamma },\mathbb{P}).$ We see this here (recall the Remark above) in
the posterior $Q_{\Gamma }$ being higher after signal $s=1$ than after
signal $s=0,$ i.e., $Q_{\Gamma }(1)=10/13>6/11=Q_{\Gamma }(0))$ together
with $\mathbb{P}^{\Gamma }$ giving a higher probability than $\mathbb{P}$ to 
$s=1$ (i.e., $\mathbb{P}^{\Gamma }(1)=5/8>13/24=\mathbb{P}(1))$. However, if
Carroll knows that the state is $\gamma ,$ then the dominance is reversed: $%
(Q_{\Gamma },\mathbb{P}^{\gamma })<_{\mathrm{lr}}(Q_{\Gamma },\mathbb{P})$
(indeed, the high-posterior signal $s=1$ then has a lower probability under $%
\mathbb{P}^{\gamma }$ than under $\mathbb{P}$ (i.e., $\mathbb{P}^{\gamma
}(1)=1/2<13/24=\mathbb{P}(1)).$ In the asset example, if Carroll knows that
the state is $\beta $\emph{\ }or $\gamma $ then he buys the asset, but if he
knows more, specifically, that it is $\gamma ,$ then he sells it.

\section{Optimism}

We provide the full proof of Proposition 3, followed by two remarks.

\bigskip

\begin{proof}[Proof of Proposition 3]
If $\widetilde{\pi }=a\pi ^{\Gamma }+(1-a)\pi $ then $\widetilde{\mathbb{P}}%
(s)=a\mathbb{P}^{\Gamma }(s)+(1-a)\mathbb{P}(s)$ for every $s,$ and so,
dividing by $\mathbb{P}(s)$ and recalling the Bayes formula (1), we have%
\[
\frac{\widetilde{\mathbb{P}}(s)}{\mathbb{P}(s)}=\frac{a}{p}Q_{\Gamma
}(s)+(1-a). 
\]%
The likelihood ratio is thus increasing in $Q_{\Gamma }(s);$ this yields the
likelihood ratio dominance as in the proof of Proposition 1.

Conversely, assume that Carroll is $\Gamma $-optimistic. Take $S=\{0,1\}$
and put $\sigma (1|\theta )=1/2+x_{\theta }$ and $\sigma (0|\theta
)=1/2-x_{\theta },$ where $|x_{\theta }|\leq 1/2.$ If $\sum_{\theta \in
\Theta }\pi (\theta )x_{\theta }=0$ and $\sum_{\theta \in \Gamma }\pi
(\theta )x_{\theta }>0$ then $\mathbb{P}(1)=\mathbb{P}(0)=1/2$ and $\mathbb{P%
}(\Gamma \cap 1)>\mathbb{P}(\Gamma \cap 0),$ and thus $Q_{\Gamma }(1)=%
\mathbb{P}(\Gamma |1)>\mathbb{P}(\Gamma |0)=Q_{\Gamma }(0).$ By $\Gamma $%
-optimism, this yields $\widetilde{\mathbb{P}}(1)\geq \mathbb{P}(1),$ i.e., $%
\sum_{\theta }\widetilde{\pi }(\theta )x_{\theta }\geq 0$ (see the Remark on
binary random variables in the previous section).

We thus have the following: if $\sum_{\theta \in \Theta }\pi (\theta
)x_{\theta }=0$ and $\sum_{\theta \in \Theta }\pi ^{\Gamma }(\theta
)x_{\theta }>0$ (we have divided the second sum by $\pi (\Gamma ))$ then $%
\sum_{\theta \in \Theta }\widetilde{\pi }(\theta )x_{\theta }\geq 0;$ this
holds for any $x=(x_{\theta })_{\theta \in \Theta }$ even if some $x_{\theta
}$ do not satisfy $|x_{\theta }|\leq 1/2$: just rescale $x$ so that they all
do, which does not affect any of the above three homogeneous relations.
Equivalently, there is no $x$ such that $\sum_{\theta \in \Theta }\pi
(\theta )x_{\theta }=0,$ $\sum_{\theta \in \Theta }\pi ^{\Gamma }(\theta
)x_{\theta }>0,$ and $\sum_{\theta \in \Theta }\widetilde{\pi }(\theta
)x_{\theta }<0,$ i.e., no $x$ such that $\pi \cdot x=0,\;\pi ^{\Gamma }\cdot
x>0,$ and $(-\widetilde{\pi })\cdot x>0.$ Motzkin's Theorem of the
Alternative (see, e.g., Mangasarian 1969) therefore yields $y,z,w$ such that 
\begin{equation}
y\pi +z\pi ^{\Gamma }+w(-\widetilde{\pi })=0  \label{eq:alt}
\end{equation}%
and $z,w\geq 0$ with at least one of them strictly positive.

For $\theta \notin \Gamma $ with $\pi (\theta )>0$ (such a $\theta $ exists
since $\pi (\Gamma )<1)$ we have $y\pi (\theta )=w\widetilde{\pi }(\theta )$
by (\ref{eq:alt}), and so $y\geq 0$ (because $w\geq 0).$ If $w=0$ then we
must have $z>0,$ but then the sum in (\ref{eq:alt}) is strictly positive for 
$\theta \in \Gamma $ with $\pi (\theta )>0$ (such a $\theta $ exists since $%
\pi (\Gamma )>0)$, a contradiction. Therefore $w>0,$ and dividing by $w$
yields%
\[
\widetilde{\pi }=a\pi ^{\Gamma }+a^{\prime }\pi , 
\]%
where $a=z/w\geq 0$ and $a^{\prime }=y/w\geq 0.$ Evaluating at $\Theta $
gives $1=a+a^{\prime },$ and the proof is complete.
\end{proof}

\bigskip

\noindent \textbf{Remarks. }\emph{(a) }The proof of the second direction
shows that when defining optimism it suffices to require first-order
stochastic dominance (instead of likelihood ratio dominance) to hold for all
binary signaling structures, i.e., $|S|=2$ (instead of all signaling
structures).

\emph{(b) }Applying Proposition 1 to $\Gamma ^{C}=\Theta \backslash \Gamma ,$
the complement of $\Gamma ,$ and using $\mathbb{P}(\Gamma ^{C}|s)=1-\mathbb{P%
}(\Gamma |s))$ yields the following: \textquotedblleft $\Gamma $-pessimism"
(which is defined by reversing the direction of the dominance relation in $%
\Gamma $-optimism) is equivalent to $\Gamma ^{C}$-strengthening (which, in
turn, is nothing other than \textquotedblleft $\Gamma $-weakening"); in
terms of the additional signal $t_{0}$ from the signaling structure $(T,\tau
)$ that Carroll receives, it means that $\tau (t_{0}|\theta )=b\leq c=\tau
(t_{0}|\theta )$ for every $\theta \in \Gamma $ and $\theta ^{\prime }\notin
\Gamma .$

\subsection{Monotonic Optimism}

We provide the full version of the statement of Proposition 4 and its proof.

Let $\pi $ and $\widetilde{\pi }$ be two prior probability distributions on%
\emph{\ }$\Theta .$ Given a signaling structure $(S,\sigma )$ on $\Theta ,$
we denote by $\mathbb{P}$ and $\widetilde{\mathbb{P}}$ the probability
distributions that are induced by the priors $\pi $ and $\widetilde{\pi },$
respectively; for each $s\in S$ and $\Gamma \subset \Theta ,$ we denote by $%
Q_{\Gamma }(s)=\mathbb{P(}\Gamma |s)$ the posterior probability of $\Gamma $
given the signal $s.$

\bigskip

\noindent \textbf{Proposition 4 }\emph{The following statements are
equivalent:}

\emph{(a) }$\widetilde{\mathbb{\pi }}$\emph{\ is a monotonic strengthening
of }$\pi $\emph{;}

\emph{(b) }$\widetilde{\pi }\in \mathrm{conv}\{\pi ^{\Gamma }:\Gamma
\subseteq \Theta $\emph{\ upper set}$\}$\emph{;}

\emph{(c) }$\widetilde{\pi }\geq _{\mathrm{lr}}\pi $\emph{\ (on }$\Theta )$%
\emph{;}

\emph{(d) }$\widetilde{\mathbb{P}}\geq _{\mathrm{lr}}\mathbb{P}$\emph{\ (on }%
$S$\emph{) for every MLRP signaling structure }$(S,\sigma )$\emph{;}

\emph{(e) }$(R,\widetilde{\mathbb{P}})\geq _{\mathrm{lr}}(R,\mathbb{P})$%
\emph{\ for every\ MLRP signaling structure }$(S,\sigma )$\emph{\ and every
increasing function\ }$R:S\rightarrow \mathbb{R}$\emph{; and}

\emph{(f) }$(Q_{\Gamma },\widetilde{\mathbb{P}})\geq _{\mathrm{lr}%
}(Q_{\Gamma },\mathbb{P})$\emph{\ for every\ MLRP signaling structure }$%
(S,\sigma )$\emph{\ and every upper set }$\Gamma $\emph{.}

\bigskip

Statements (a) and (f) provide the main equivalence of Proposition 4 in the
paper; (c), (d), and (e) here are, respectively, (i), (ii), and (iii) there.

\bigskip

\begin{proof}
The proof contains two cycles: first, (a) $\Rightarrow $ (c) $\Rightarrow $
(b) $\Rightarrow $ (a), and then (c) $\Rightarrow $ (d) $\Rightarrow $ (e) $%
\Rightarrow $ (f) $\Rightarrow $ (c).

Monotonicity of ratios, i.e., $a^{\prime }/b^{\prime }\geq a/b,$ is always
taken to mean $a^{\prime }b\geq ab^{\prime },$ which applies also when
denominators equal zero. We assume without loss of generality that $\pi
(\theta )>0$ for all $\theta \in \Theta .$

\emph{(a) implies (c).} If $\pi _{2}=a\pi _{1}^{\Gamma }+(1-a)\pi _{1}$ is a 
$\Gamma $-strengthening of $\pi _{1}$ then 
\[
\frac{\pi _{2}(\theta )}{\pi _{1}(\theta )}=\frac{a}{\pi _{1}(\Gamma )}%
\mathbf{1}_{\Gamma }(\theta )+1-a, 
\]%
where $\mathbf{1}_{\Gamma }$ denotes the indicator of the set $\Gamma .$
When $\Gamma $ is an upper set this ratio is increasing in $\theta ,$ and
thus $\pi _{2}\geq _{\mathrm{lr}}\pi _{1}.$ Using the transitivity of the
likelihood ratio dominance (because $\pi _{3}/\pi _{1}=(\pi _{3}/\pi
_{2})\cdot (\pi _{2}/\pi _{1}))$ completes the proof.

\emph{(c) implies (b). }Let $\theta _{1}<\theta _{2}<...<\theta _{n}$ be the
elements of $\Theta .$ If $\widetilde{\pi }\geq _{\mathrm{lr}}\pi $ then the
increasing function $h(\theta ):=\widetilde{\pi }(\theta )/\pi (\theta )$
can be represented as $\sum_{i=1}^{n}a_{i}\mathbf{1}_{\Gamma _{i}}(\theta )$
where $\Gamma _{i}:=\{\theta _{i},...,\theta _{n}\}$ (an upper set) and $%
a_{i}\geq 0$ for all $i.$ Therefore $\widetilde{\pi }(\theta
)=\sum_{i=1}^{n}a_{i}\mathbf{1}_{\Gamma _{i}}(\theta )\pi (\theta
)=\sum_{i=1}^{n}a_{i}\pi (\Gamma _{i})\pi ^{\Gamma _{i}}(\theta ),$ showing
that $\widetilde{\pi }$ is a convex combination of the $\pi ^{\Gamma _{i}}$
(summing over all $\theta $ shows that the coefficients add up to $1,$
because $\widetilde{\pi }$ and the $\pi ^{\Gamma _{i}}$ are all probability
distributions).

\emph{(b) implies (a).} Take $\pi _{0}=\sum_{i=1}^{m}a_{i}\pi ^{\Gamma
_{i}}, $ where $a_{i}\geq 0$, $\sum_{i=1}^{m}a_{i}=1,$ and the $\Gamma _{i}$
are in decreasing order, i.e., $\Gamma _{1}\supset \Gamma _{2}\supset
...\supset \Gamma _{m}.$ Then 
\begin{equation}
\pi _{0}=a_{m}\pi ^{\Gamma _{m}}+(1-a_{m})\pi _{1},  \label{eq:pi0}
\end{equation}%
where $\pi _{1}=\sum_{i=1}^{m-1}b_{i}\pi ^{\Gamma _{i}}$ and $%
b_{i}=a_{i}/(1-a_{m}).$ Now%
\begin{eqnarray*}
\pi _{1}^{\Gamma _{m}}(\theta ) &=&\frac{1}{\pi _{1}(\Gamma _{m})}%
\sum_{i=1}^{m-1}b_{i}\pi ^{\Gamma _{i}}(\theta \cap \Gamma ^{m})=\frac{1}{%
\pi _{1}(\Gamma _{m})}\sum_{i=1}^{m-1}b_{i}\frac{\pi (\Gamma _{m})}{\pi
(\Gamma _{i})}\pi ^{\Gamma _{m}}(\theta ) \\
&=&\left( \sum_{i=1}^{m-1}c_{i}\right) \pi ^{\Gamma _{m}}(\theta )=\pi
^{\Gamma _{m}}(\theta ),
\end{eqnarray*}%
by $\Gamma _{i}\supset \Gamma _{m}$ for $i=1,...,m-1,$ and $%
\sum_{i=1}^{m-1}c_{i}=1$ since $\pi _{1}^{\Gamma _{m}}$ and $\pi ^{\Gamma
_{m}}$ are both probability measures. Substituting in (\ref{eq:pi0}) gives 
\[
\pi _{0}=a_{m}\pi _{1}^{\Gamma _{m}}+(1-a_{m})\pi _{1}, 
\]%
and so $\pi _{0}$ is a $\Gamma _{m}$-strengthening of $\pi _{1}.$ Induction
on $m$ completes the proof.

\emph{(c) implies (d). }This is shown by a standard composition argument ($%
\mathbb{P}$ and $\widetilde{\mathbb{P}}$ are the compositions of $\sigma $
with $\pi $ and $\widetilde{\mathbb{\pi }},$ respectively; see, e.g., Karlin
1968, formulas (0.1) and (1.1), which are versions of the Cauchy--Binet
formula). We provide here (and in the following arguments as well) a direct
proof. We have 
\[
\widetilde{\mathbb{P}}(s^{\prime })\mathbb{P}(s)-\widetilde{\mathbb{P}}(s)%
\mathbb{P}(s^{\prime })=\sum \sum [\widetilde{\pi }(\theta ^{\prime })\pi
(\theta )-\widetilde{\pi }(\theta )\pi (\theta ^{\prime })]\cdot \lbrack
\sigma (s^{\prime }|\theta ^{\prime })\sigma (s|\theta )-\sigma (s|\theta
^{\prime })\sigma (s^{\prime }|\theta )], 
\]%
where the double sum is over all $\theta ^{\prime }>\theta $. When $%
s^{\prime }>s$ the terms on the right-hand side are all $\geq 0$ by $%
\widetilde{\pi }\geq _{\mathrm{lr}}\pi $ and MLRP.

\emph{(d) implies (e).} We have%
\[
\widetilde{\mathbb{P}}(R=r^{\prime })\mathbb{P}(R=r)-\widetilde{\mathbb{P}}%
(R=r)\mathbb{P}(R=r^{\prime })=\sum_{\QATOP{s^{\prime }\in S:}{R(s^{\prime
})=r^{\prime }}}\sum_{\QATOP{s\in S:}{R(s)=r}}[\widetilde{\mathbb{P}}%
(s^{\prime })\mathbb{P}(s)-\widetilde{\mathbb{P}}(s)\mathbb{P}(s^{\prime
})]. 
\]%
When $r^{\prime }>r$ the terms on the right-hand are all $\geq 0$ because $%
R(s^{\prime })=r^{\prime }>r=R(s)$ implies $s^{\prime }\geq s.$

\emph{(e) implies (f).} We need to show that $Q_{\Gamma }$ is increasing for
every upper set\footnote{%
The functions $Q_{\Gamma }$ for all upper sets $\Gamma $ are thus
\textquotedblleft co-monotone."} $\Gamma .$ We have 
\begin{eqnarray*}
\mathbb{P}(\Gamma \cap s^{\prime })\mathbb{P}(s)-\mathbb{P}(\Gamma \cap s)%
\mathbb{P}(s^{\prime }) &=&\mathbb{P}(\Gamma \cap s^{\prime })\mathbb{P}%
(\Gamma ^{C}\cap s)-\mathbb{P}(\Gamma \cap s)\mathbb{P}(\Gamma ^{C}\cap
s^{\prime }) \\
&=&\sum_{\theta ^{\prime }\in \Gamma }\sum_{\theta \in \Gamma ^{C}}\pi
(\theta ^{\prime })\pi (\theta )\left[ \sigma (s^{\prime }|\theta ^{\prime
})\sigma (s|\theta )-\sigma (s^{\prime }|\theta )\sigma (s|\theta ^{\prime })%
\right] .
\end{eqnarray*}%
When $s^{\prime }>s$ the terms on the right-hand side are $\geq 0$ by MLRP,
because $\theta ^{\prime }\in \Gamma $ and $\theta \in \Gamma ^{C}$ implies $%
\theta ^{\prime }>\theta $ (since $\Gamma $ is an upper set).

\emph{(f) implies (c). }Let $\theta _{1}<\theta _{2}$ be two adjacent
elements of $\Theta $ (i.e., there is no $\theta \in \Theta $ between $%
\theta _{1}$ and $\theta _{2}),$ and take the following signaling structure: 
$S=\{0,1,2,3\};$ for $\theta <\theta _{1}$ the signal is $s=0,$ for $\theta
>\theta _{2}$ it is $s=3,$ for $\theta =\theta _{1}$ it is $s=1$ with
probability $2/3$ and $s=2$ with probability $1/3,$ and for $\theta =\theta
_{2}$ it is $s=1$ with probability $1/3$ and $s=2$ with probability $2/3.$
Clearly, $(S,\sigma )$ is MLRP. Take the upper set $\Gamma =\{\theta \in
\Theta :\theta \geq \theta _{2}\}.$ The posterior probabilities of $\Gamma $
are $Q_{\Gamma }(0)=0<Q_{\Gamma }(1)<Q_{\Gamma }(2)<1=$ $Q_{\Gamma }(3),$
where%
\begin{eqnarray*}
Q_{\Gamma }(1) &=&\frac{\frac{1}{3}\pi (\theta _{2})}{\frac{2}{3}\pi (\theta
_{1})+\frac{1}{3}\pi (\theta _{2})}=\frac{1}{2\frac{\pi (\theta _{1})}{\pi
(\theta _{2})}+1}\text{\ \ and} \\
Q_{\Gamma }(2) &=&\frac{\frac{2}{3}\pi (\theta _{2})}{\frac{1}{3}\pi (\theta
_{1})+\frac{2}{3}\pi (\theta _{2})}=\frac{1}{\frac{1}{2}\frac{\pi (\theta
_{1})}{\pi (\theta _{2})}+1}.
\end{eqnarray*}%
Therefore%
\[
\frac{\frac{1}{3}\widetilde{\pi }(\theta _{1})+\frac{2}{3}\widetilde{\pi }%
(\theta _{2})}{\frac{1}{3}\pi (\theta _{1})+\frac{2}{3}\pi (\theta _{2})}=%
\frac{\widetilde{\mathbb{P}}(Q_{\Gamma }=\alpha _{2})}{\mathbb{P}(Q_{\Gamma
}=\alpha _{2})}\geq \frac{\widetilde{\mathbb{P}}(Q_{\Gamma }=\alpha _{1})}{%
\mathbb{P}(Q_{\Gamma }=\alpha _{1})}=\frac{\frac{2}{3}\widetilde{\pi }%
(\theta _{1})+\frac{1}{3}\widetilde{\pi }(\theta _{2})}{\frac{2}{3}\pi
(\theta _{1})+\frac{1}{3}\pi (\theta _{2})}, 
\]%
where the inequality is by (f). Simplifying yields $\widetilde{\pi }(\theta
_{2})/\pi (\theta _{2})\geq \widetilde{\pi }(\theta _{1})/\pi (\theta _{1});$
since this holds for any adjacent $\theta _{1}<\theta _{2},$ we have $%
\widetilde{\pi }\geq _{\mathrm{lr}}\pi $.
\end{proof}

%TCIMACRO{%
%\TeXButton{References Without Numbers}{\def\@biblabel#1{#1\hfill}
%\def\thebibliography#1{\section*{References}
%\addcontentsline{toc}{section}{References}
%\list
%{}{
%\labelwidth 0pt
%\leftmargin 1.8em
%\itemindent -1.8em
%\usecounter{enumi}}
%\def\newblock{\hskip .11em plus .33em minus .07em}
%\sloppy\clubpenalty4000\widowpenalty4000
%\sfcode`\.=1000\relax\def\baselinestretch{1}\large \normalsize}
%\let\endthebibliography=\endlist}}%
%BeginExpansion
\def\@biblabel#1{#1\hfill}
\def\thebibliography#1{\section*{References}
\addcontentsline{toc}{section}{References}
\list
{}{
\labelwidth 0pt
\leftmargin 1.8em
\itemindent -1.8em
\usecounter{enumi}}
\def\newblock{\hskip .11em plus .33em minus .07em}
\sloppy\clubpenalty4000\widowpenalty4000
\sfcode`\.=1000\relax\def\baselinestretch{1}\large \normalsize}
\let\endthebibliography=\endlist%
%EndExpansion

\end{document}

%TCIMACRO{%
%\TeXButton{References Without Numbers}{\def\@biblabel#1{#1\hfill}
%\def\thebibliography#1{\section*{References}
%\addcontentsline{toc}{section}{References}
%\list
%{}{
%\labelwidth 0pt
%\leftmargin 1.8em
%\itemindent -1.8em
%\usecounter{enumi}}
%\def\newblock{\hskip .11em plus .33em minus .07em}
%\sloppy\clubpenalty4000\widowpenalty4000
%\sfcode`\.=1000\relax\def\baselinestretch{1}\large \normalsize}
%\let\endthebibliography=\endlist}}%
%BeginExpansion
\def\@biblabel#1{#1\hfill}
\def\thebibliography#1{\section*{References}
\addcontentsline{toc}{section}{References}
\list
{}{
\labelwidth 0pt
\leftmargin 1.8em
\itemindent -1.8em
\usecounter{enumi}}
\def\newblock{\hskip .11em plus .33em minus .07em}
\sloppy\clubpenalty4000\widowpenalty4000
\sfcode`\.=1000\relax\def\baselinestretch{1}\large \normalsize}
\let\endthebibliography=\endlist%
%EndExpansion

Alice is a classical Bayesian decision-maker, in a setup with a state that
is unknown, a prior probability distribution over the states, and signals
whose distribution depends on the state. Alice is interested in a particular
\textquotedblleft good" state $\gamma ,$ and so she computes, by Bayes'
rule, the posterior probability of $\gamma $ for each possible signal; call
this posterior probability $Q_{\gamma }.$ Before getting the signal, Alice
thus believes that $Q_{\gamma }$ will be either higher or lower than the
prior probability $p$ of $\gamma $, with an expectation exactly equal to $p.$

Now suppose that Carroll (an outside observer, say) knows that the state is
in fact $\gamma ,$ and he considers what will be the posterior probability $%
Q_{\gamma }$ computed by Alice (who does not know that the state is $\gamma
).$ How does Carroll's (probabilistic) belief on $Q_{\gamma }$ compare to
Alice's belief (both beliefs are of course before the signal)?\footnote{%
Lest there be any misunderstanding, we are \emph{not} considering Carroll's
posterior probability of $\gamma $ (which equals $1),$ but rather Carroll's
view of Alice's posterior probability $Q_{\gamma }$ of $\gamma .$ For simple
setups where this is a natural question, e.g., Alice is the uninformed
market and Carroll is an informed agent, see Section \ref{s:dominance} below.%
} It turns out that Carroll always assigns higher probability than Alice to
high values of $Q_{\gamma },$ and lower probability to low values; formally,
we will show that Carroll's probability distribution of $Q_{\gamma }$
stochastically dominates Alice's in the so-called \textquotedblleft
likelihood ratio" order, which is stronger than first-order stochastic
dominance.\footnote{%
A standard reference on stochastic orders is the book of Shaked and
Shanthikumar (2010).} What may come as a surprise here is that this fact,
whose proof is straightforward, is unknown. Even its immediate consequence
that Carroll's expectation of $Q_{\gamma }$ is higher than the prior $p$
(which is equal to Alice's expectation of $Q_{\gamma })$ seems to be still
widely unknown despite having appeared in the literature.\footnote{\label%
{ftn:refs}Mailath and Samuelson (2006); Francetich and Kreps (2014). None of
the colleagues whom we asked knew this result, just as those whom Francetich
and Kreps asked didn't.}

The Supplementary Material\footnote{%
Available at \texttt{http://www.ma.huji.ac.il/hart/abs/posterior.html}}
contains full proofs and additional comments.

\section{Dominance\label{s:dominance}}

The Bayesian setup that Alice faces is standard; for simplicity, all sets
are assumed to be finite. The set of \emph{states of nature} is $\Theta ,$
and $\pi $ is the \emph{prior} probability distribution on $\Theta $. There
is a set $\Gamma \subset \Theta $ of states of nature that is of interest; $%
\Gamma $ could well consist of a single state $\gamma .$ To avoid
trivialities, assume that its prior probability $p\equiv \pi (\Gamma )$
satisfies $0<p<1.$ Let $S$ be the set of \emph{signals} and $\sigma
(s|\theta )$ the probability of signal $s$ in state $\theta $ (thus $%
\sum_{s\in S}\sigma (s|\theta )=1$ for all $\theta );$ we will refer to $%
(S,\sigma )$ as a \emph{signaling structure} (on $\Theta $). The prior
probability $\pi $ and the signaling probabilities $\sigma $ induce a
probability $\mathbb{P\equiv \mathbb{P}}_{\pi ,\sigma }$ on states and
signals, i.e., on $\Theta \times S$.

For each signal $s$ in $S,$ Alice's \emph{posterior} probability of $\Gamma
, $ which we denote by $Q_{\Gamma }(s),$ is given by Bayes' formula:

\begin{equation}
Q_{\Gamma }(s)\equiv \mathbb{P}(\Gamma |s)=\frac{\mathbb{P}(\Gamma )\cdot 
\mathbb{P}(s|\Gamma )}{\mathbb{P}(s)}=p\cdot \frac{\mathbb{P}^{\Gamma }(s)}{%
\mathbb{P}(s)},  \label{eq:bayes}
\end{equation}%
where $\mathbb{P}^{\Gamma }$ denotes the conditional-on-$\Gamma $
probability, i.e.,\footnote{%
Superscripts (on $\mathbb{P},$ $\mathbb{E},$ $\pi )$ are used throughout to
denote conditioning. When $\Gamma $ consists of a single state $\gamma $ we
have $\mathbb{P}^{\gamma }(s)=\sigma (s|\gamma ).$} $\mathbb{P}^{\Gamma
}(s)\equiv \mathbb{P}(s|\Gamma )=$\linebreak $(1/p)\sum_{\theta \in \Gamma
}\pi (\theta )\sigma (s|\theta ),$ and $\mathbb{P}(s)=\sum_{\theta \in
\Theta }\pi (\theta )\sigma (s|\theta )$. Alice's posterior of $\Gamma $ is
a random variable that takes the value $Q_{\Gamma }(s)$ with probability $%
\mathbb{P}(s);$ its expectation equals the prior, i.e., $\mathbb{\mathbb{E}}%
\left[ Q_{\Gamma }\right] =\pi (\Gamma )=p.$

Carroll knows that the state is in fact in $\Gamma .$ He thus assigns
probability $\mathbb{P}^{\Gamma }(s)$ to the signal $s,$ and so he believes
that Alice's posterior of $\Gamma $ takes the value $Q_{\Gamma }(s)$ with
probability $\mathbb{P}^{\Gamma }(s)$ (instead of $\mathbb{P}(s)).$ We
denote this random variable by $(Q_{\Gamma },\mathbb{P}^{\Gamma }),$ whereas
Alice's posterior random variable, which takes the value $Q_{\Gamma }(s)$
with probability $\mathbb{P}(s),$ is denoted by $(Q_{\Gamma },\mathbb{P}).$

What is the relation between $(Q_{\Gamma },\mathbb{P}^{\Gamma })$ and $%
(Q_{\Gamma },\mathbb{P})$ ? It turns out that high values of the posterior $%
Q_{\Gamma }$ are more probable under $\mathbb{P}^{\Gamma }$ than under $%
\mathbb{P},$ whereas low values of the posterior are more probable under $%
\mathbb{P}$ than under $\mathbb{P}^{\Gamma };$ formally, the former is
always higher\footnote{%
All comparisons such as \textquotedblleft higher" and \textquotedblleft
increasing" are meant in the weak-inequality sense.} than the latter, in the
following precise sense.

\begin{proposition}
\label{p:LR}$(Q_{\Gamma },\mathbb{P}^{\Gamma })$ dominates $(Q_{\Gamma },%
\mathbb{P})$ in the likelihood ratio order.
\end{proposition}

The likelihood ratio stochastic order is stronger than first-order
stochastic dominance. Formally, let $X$ and $Y$ be two random variables.
Then $X$ \emph{first-order stochastically dominates} $Y$ if $\mathbb{P}%
(X\geq v)\geq \mathbb{P}(Y\geq v)$ for every $v$; and $X$ \emph{dominates} $%
Y $ \emph{in the likelihood ratio order}, denoted by $X\geq _{\mathrm{lr}}Y,$
if $\mathbb{P}(X=v)/\mathbb{P(}Y=v)$ is an increasing function of $v.$ The
former is equivalent to $\mathbb{E}\left[ \varphi (X)\right] \geq \mathbb{E}%
\left[ \varphi (Y)\right] $ for every increasing function $\varphi ,$ and
the latter to $\mathbb{E}\left[ \varphi (X)|X\in A\right] \geq \mathbb{E}%
\left[ \varphi (Y)|Y\in A\right] $ for every increasing function $\varphi $
and every measurable set of values $A\subseteq \mathbb{R}.$

\bigskip

\begin{proof}
Immediate, since $Q_{\Gamma }$ is proportional to the likelihood ratio $%
\mathbb{P}^{\Gamma }/\mathbb{P}$ by Bayes' formula (\ref{eq:bayes}).
Formally, $Q_{\Gamma }(s)=q$ if and only if $\mathbb{P}^{\Gamma
}(s)=(q/p)\cdot \mathbb{P}(s),$ and so summing over all $s$ with $Q_{\Gamma
}(s)=q$ yields $\mathbb{P}^{\Gamma }(Q_{\Gamma }=q)=(q/p)\cdot \mathbb{P}%
(Q_{\Gamma }=q).$ The likelihood ratio thus equals $q/p,$ which is an
increasing function of $q.$
\end{proof}

\bigskip 

The dominance is in fact \emph{strict} (i.e., the likelihood ratio is a
strictly increasing function of $Q_{\Gamma }$), except when the signal is
completely uninformative about $\Gamma $ (i.e., for every signal the
posterior of $\Gamma $ equals its prior $p$).

Despite its simplicity, this result appears not to be known. One easy
implication that does appear in the literature (see footnote \ref{ftn:refs})
is the resulting inequality on expectations:

\begin{corollary}
\label{c:submart} $\mathbb{\mathbb{E}}^{\Gamma }\left[ Q_{\Gamma }\right]
\geq \mathbb{E}\left[ Q_{\Gamma }\right] =\pi (\Gamma ).$
\end{corollary}

Thus, the conditional-on-$\Gamma $ expectation of the posterior of $\Gamma $
is higher than the prior of $\Gamma ;$ that is, the posterior probability of 
$\Gamma $ is a submartingale relative to $\mathbb{P}^{\Gamma }$. Moreover,
the inequality is strict except when the signal is completely uninformative
about $\Gamma $.

When does this stochastic domination matter? For a simple setup, consider an
uncertain asset whose market price is determined by the estimated
probability that the state is in a set of \textquotedblleft good" states $%
\Gamma ;$ specifically, the price is an increasing function $\varphi $ of
the probability of $\Gamma .$ Suppose that a signal whose distribution
depends on the state (such as a quarterly report or a management
announcement) is forthcoming. Before the signal the expectation of the price
according to the uninformed market (represented by Alice) is thus $\mathbb{E}%
\left[ \varphi (Q_{\Gamma })\right] ,$ whereas for an informed trader,
Carroll, who knows that the state is in fact in $\Gamma ,$ this expectation
is $\mathbb{E}^{\Gamma }\left[ \varphi (Q_{\Gamma })\right] .$ Our result
says that Carroll's expectation of the price is always higher than that of
the market. Moreover, this holds even if one conditions on a certain set of
values of the posterior, such as the posterior being within certain bounds: $%
\mathbb{E}^{\Gamma }\left[ \varphi (Q_{\Gamma })|Q_{\Gamma }\in A\right]
\geq \mathbb{E}\left[ \varphi (Q_{\Gamma })|Q_{\Gamma }\in A\right] $ for
every $A\subseteq \lbrack 0,1]$. Thus, no matter what the information
structure is, the informed trader who knows that the state is in $\Gamma $
is ex-ante (i.e., before the signal) more \textquotedblleft optimistic" than
the market about the ex-post (i.e., after the signal) price.\footnote{%
We emphasize that the optimism is not about $\Gamma $ per se (which goes
without saying), but rather about the distribution of the market's posterior
of $\Gamma $ after the signal, and thus about the market's behavior then.}
He will therefore buy the asset now.

A general setup consists of players that possess different information (such
as uninformed players, like Alice, and informed players, like Carroll), with
actions that depend monotonically on beliefs (such as \textquotedblleft
threshold strategies"), and these actions may be unobserved or imperfectly
monitored. One specific such class is that of reputation models with
imperfect monitoring (see Fudenberg and Levine 1992, and Mailath and
Samuelson 2006, Section 15.4): a sequence of short-lived players P2 face a
long-lived player P1, who may be of a type $\gamma $ that is committed to
playing a fixed action. If the non-committed type of P1 plays like the
committed type $\gamma $ then P2's posterior belief that P1 \emph{is} of
type $\gamma $ becomes higher in the likelihood ratio order,\footnote{%
Lemma 15.4.2 in Mailath and Samuelson (2006) gives this in expectation
(i.e., Corollary \ref{c:submart}).} thus strengthening P1's
\textquotedblleft reputation," and potentially increasing his payoff.

In Bayesian statistics, where $\theta $ is an unknown parameter with prior
distribution $\pi ,$ the interpretation of our results for, say, $\Gamma
=\{\theta _{0}\}$ is that data generated under $\theta _{0}$ (i.e., an
observation $s$ that is distributed according to $\mathbb{P}^{\theta _{0}})$
increases a Bayesian's belief in $\theta _{0}$ in the likelihood order
sense, and thus also in expectation: $\mathbb{E}^{\theta _{0}}[\mathbb{P}%
(\theta _{0}|s)]\geq \pi (\theta _{0}).$ See Hart and Rinott (2020) for a
study of sequences of observations.

Finally, a word of caution. If Carroll knows \emph{more} than $\Gamma $%
---for instance, if $\Gamma =\{\beta ,\gamma \}$ and Carroll knows that the
state is in fact $\gamma $---then some of these inequalities may well be
reversed: while knowing $\Gamma $ makes one optimistic (relative to Alice,
or the market) about the posterior of $\Gamma ,$ knowing strictly more than $%
\Gamma $ may well turn an optimist into a pessimist. For a simple example,
suppose that the set of states is $\Theta =\{\alpha ,\beta ,\gamma \},$ the
prior $\pi $ is uniform on $\Theta $ (i.e., each state has probability $%
1/3), $ and $\Gamma =\{\beta ,\gamma \}.$ Let $S=\{0,1\}$ be the set of
signals, with signaling probabilities $\sigma (s|\theta )$ as given in the
table below (the last two columns then give the total probability $\mathbb{P}%
(s)$ of each signal $s$ and the posterior $Q_{\Gamma }(s)\equiv \mathbb{P}%
(\Gamma |s))$:%
\[
{\renewcommand{\arraystretch}{2.8}\setlength{\tabcolsep}{7pt}}%
\begin{tabular}{c||ccc||c|c||c}
$s$ & $\sigma (s|\alpha )$ & $\mathbb{\sigma }(s|\beta )$ & $\mathbb{\sigma }%
(s|\gamma )$ & $\mathbb{P}(s)$ & $\mathbb{P}^{\Gamma }(s)$ & $Q_{\Gamma }(s)$
\\ \hline\hline
$1$ & $3/8$ & $3/4$ & $1/2$ & $13/24$ & $5/8$ & $10/13$ \\ 
$0$ & $5/8$ & $1/4$ & $1/2$ & $11/24$ & $3/8$ & $6/11$%
\end{tabular}%
\]%
We have $(Q_{\Gamma },\mathbb{P}^{\Gamma })>_{\mathrm{lr}}(Q_{\Gamma },%
\mathbb{P})$ (by Proposition \ref{p:LR}). However, if Carroll knows that the
state is $\gamma ,$ then the dominance is reversed: $(Q_{\Gamma },\mathbb{P}%
^{\gamma })<_{\mathrm{lr}}(Q_{\Gamma },\mathbb{P}).$ In the asset example,
if Carroll knows that the state is $\beta $\emph{\ }or $\gamma $ then he
buys the asset, but if he knows more, specifically, that it is $\gamma ,$
then he sells it.

\section{Optimism\label{s:optimism}}

The result of the previous section suggests the following question: under
what circumstances---besides knowing that the state is in $\Gamma $---is
Carroll's belief about $Q_{\Gamma }$ \emph{guaranteed} to be higher than
Alice's belief, no matter what the signaling structure is? Suppose that
Carroll has a prior distribution $\widetilde{\pi }$ over the states of
nature in $\Theta $ that may be different from Alice's prior $\pi $ (in the
previous section, $\widetilde{\pi }$ is $\pi ^{\Gamma }).$ Let $\widetilde{%
\mathbb{P}}\equiv \mathbb{P}_{\widetilde{\pi },\sigma }$ denote Carroll's
probability on $\Theta \times S$ that is induced by $\widetilde{\pi }$ and $%
\sigma $. Before the signal $s,$ Carroll believes that Alice's posterior
probability of $\Gamma $ will be $Q_{\Gamma }(s)$ with probability $%
\widetilde{\mathbb{P}}(s);$ we denote this random variable by $(Q_{\Gamma },%
\widetilde{\mathbb{P}}).$

We introduce two concepts that compare posteriors and priors, respectively.

$\bullet $ Carroll is $\Gamma $-\emph{optimistic }if $\mathbb{(}Q_{\Gamma },%
\widetilde{\mathbb{P}})$ dominates $(Q_{\Gamma },\mathbb{P})$ in the
likelihood ratio order for every signaling structure $(S,\sigma )$ on the
state space $\Theta .$

$\bullet $ Carroll's prior $\widetilde{\pi }$ is a $\Gamma $\emph{%
-strengthening} of the prior $\pi $ if $\widetilde{\pi }$ is an average of $%
\pi $ and $\pi ^{\Gamma },$ i.e., $\widetilde{\pi }=a\pi ^{\Gamma }+(1-a)\pi 
$ for some $a\in \lbrack 0,1]$ (where $\pi ^{\Gamma }(\theta )\equiv \pi
(\theta |\Gamma )$ is the prior probability of $\theta $ conditional on $%
\Gamma ).$

It is readily seen that $\Gamma $-strengthening is equivalent to starting
with the same prior $\pi $ and then receiving (before the signal $S)$ a
specific signal $t_{0}\in T$ from another signaling structure $(T,\tau )$
that satisfies $\mathbb{\tau }(t_{0}|\theta )=b\geq c=\tau (t_{0}|\theta
^{\prime })$ for every $\theta \in \Gamma $ and $\theta ^{\prime }\notin
\Gamma $. A special case is where Carroll knows exactly $\Gamma ,$ i.e., $%
\widetilde{\pi }=\pi ^{\Gamma }.$

The two concepts turn out to be equivalent.

\begin{proposition}
\label{p:Gamma}Carroll is $\Gamma $-optimistic if and only if Carroll's
prior $\widetilde{\pi }$ is a $\Gamma $-strengthening of the prior $\pi .$
\end{proposition}

\begin{proof}[Proof outline]
$\Gamma $-strengthening implies that the likelihood ratio $\widetilde{%
\mathbb{P}}/\mathbb{P}$ equals $(a/p)Q_{\Gamma }+1-a,$ which is monotonic in 
$Q_{\Gamma },$ and so $\Gamma $-optimism follows. For the converse take a
signaling structure with $S=\{0,1\},$ $\sigma (0|\theta )=1/2-x_{\theta },$ $%
\sigma (1|\theta )=1/2+x_{\theta },$ $\sum_{\theta \in \Theta }\pi (\theta
)x_{\theta }=0,$ and $\sum_{\theta \in \Gamma }\pi (\theta )x_{\theta }>0;$
by $\Gamma $-optimism, we have $\sum_{\theta \in \Theta }\widetilde{\pi }%
(\theta )x_{\theta }\geq 0.$ Thus, $\pi \cdot x=0$ and $\pi ^{\Gamma }\cdot
x>0$ imply $\widetilde{\pi }\cdot x\geq 0.$ Using a standard theorem of the
alternative we then show that $\widetilde{\pi }=a\pi ^{\Gamma }+b\pi $ for
some $a,b\geq 0;$ evaluating at $\Theta $ yields $1=a+b.$
\end{proof}

\subsection{Monotonic Optimism\label{s:monot-optimism}}

Suppose now that there is more than one state (or one set of states) that
matters. For example, the price of the asset may well be determined by the
entire probability distribution on $\Theta $ that is the market's belief,
and not just by the market's probability of one specific set $\Gamma $ of
\textquotedblleft good" states. What does it take to be optimistic in this
case? Unfortunately, Proposition \ref{p:Gamma} implies that already for two
distinct sets $\Gamma _{1}$ and $\Gamma _{2}$, Carroll cannot be both $%
\Gamma _{1}$-optimistic and $\Gamma _{2}$-optimistic unless $\widetilde{\pi }%
=\pi $ (that is, Carroll has the same information as Alice).

The definition of optimism requires taking into account \emph{all} possible
signaling structures, which suggests restricting them to a natural subclass
of interest. We consider the commonly used class of signaling structures
that have the \emph{monotone likelihood ratio property (MLRP)}; i.e., $%
\sigma (s^{\prime }|\theta )/\sigma (s|\theta )$ is increasing in $\theta $
for $s^{\prime }>s,$ where $\Theta $ and $S$ are subsets of, say, the real
line. The interpretation is that high $\theta $'s represent
\textquotedblleft good" states and low $\theta $'s, \textquotedblleft bad"
states, and the higher a signal $s$ is, the more indicative $s$ is of higher
states $\theta $. In such setups one considers \emph{upper} sets of states $%
\Gamma \subseteq \Theta ,$ for which $\theta \in \Gamma $ implies $\theta
^{\prime }\in \Gamma $ for all higher $\theta ^{\prime }>\theta $ in $\Theta
.$ Returning to the asset example, it is only natural for its price $R$ to
increase as the probability of the good states increases (more precisely, as
probability mass is moved from low $\theta $ to high $\theta $). This means
that $R,$ as a function of probability distributions on the state space $%
\Theta ,$ is \emph{increasing with respect to (first-order) stochastic
dominance}; that is,\footnote{%
Since $\pi _{1}$ first-order stochastically dominates $\pi _{2}$ if and only
if $\pi _{1}(\Gamma )\geq \pi _{2}(\Gamma )$ for all upper sets $\Gamma .$} $%
R=\varphi (Q_{\Gamma _{1}},Q_{\Gamma _{2}},...,Q_{\Gamma _{m}})$ with $%
\varphi $ increasing in each coordinate and all the $\Gamma _{i}$ upper sets.

We thus define the monotonic versions of our two concepts:

$\bullet $ Carroll is \emph{monotonic-optimistic} if $(Q_{\Gamma },%
\widetilde{\mathbb{P}})\geq _{\mathrm{lr}}(Q_{\Gamma },\mathbb{P})$ for
every upper set $\Gamma \subseteq \Theta $ and every MLRP signaling
structure $(S,\sigma )$ on the state space $\Theta .$

$\bullet $ Carroll's prior $\widetilde{\pi }$ is a \emph{monotonic
strengthening} of the prior $\pi $ if $\widetilde{\pi }$ is obtained by a
finite sequence of $\Gamma _{i}$-strengthenings that starts from $\pi $ and
all the $\Gamma _{i}\subset \Theta $ are upper sets.

\begin{proposition}
\label{p:upper}Carroll is monotonic-optimistic if and only if Carroll's
prior $\widetilde{\mathbb{\pi }}$ is a monotonic strengthening of the prior $%
\pi .$

Moreover, these are also equivalent to each one of the following statements:

(i) $\widetilde{\pi }\geq _{\mathrm{lr}}\pi $ (on $\Theta $);

(ii) $\widetilde{\mathbb{P}}\geq _{\mathrm{lr}}\mathbb{P}$ (on $S$) for
every MLRP signaling structure $(S,\sigma )$; and

(iii) $(R,\widetilde{\mathbb{P}})\geq _{\mathrm{lr}}(R,\mathbb{P})$ for every%
\emph{\ }MLRP signaling structure $(S,\sigma )$ and every increasing function%
\emph{\ }$R:S\rightarrow \mathbb{R}.$
\end{proposition}

MLRP implies that higher signals make upper sets more probable, and so $%
Q_{\Gamma }(s)$ increases in $s$ for every upper set $\Gamma .$ Therefore,
the functions $R$ in condition (iii) include all these $Q_{\Gamma },$ as
well as all their increasing functions, as discussed above. Monotonicity
with respect to stochastic dominance is thus equivalent, by Proposition \ref%
{p:upper}, to the concept of monotonic optimism that we defined (as optimism
for upper sets).

\bigskip 

\begin{proof}[Proof outline]
First, monotonic strengthening is equivalent to (i): for one direction,
every upper-set strengthening yields $\widetilde{\pi }\geq _{\mathrm{lr}}\pi
,$ and $\geq _{\mathrm{lr}}$ is transitive; for the converse, express the
increasing function $\widetilde{\pi }/\pi $ as a positive linear combination
of indicators of upper sets, which translates to $\widetilde{\pi }$ being a
convex combination of $\pi ^{\Gamma }$'s for upper sets $\Gamma $, and from
which we then obtain a sequence of upper-set strengthenings from $\pi $ to $%
\widetilde{\pi }.$

Second, standard composition arguments show that (i) implies (ii); (ii)
implies (iii); and (iii) includes monotonic optimism (as we saw above).

Finally, monotonic optimism implies (i): for any two adjacent elements $%
\theta _{1}<\theta _{2}$ of $\Theta $ we construct an MLRP $(S,\sigma )$ and
an upper set $\Gamma $ such that $(Q_{\Gamma },\widetilde{\mathbb{P}})\geq _{%
\mathrm{lr}}(Q_{\Gamma },\mathbb{P})$ yields $\widetilde{\pi }(\theta
_{2})/\pi (\theta _{2})\geq \widetilde{\pi }(\theta _{1})/\pi (\theta _{1}).$
\end{proof}

\bigskip 

Suppose that there is another agent, Bob, and that Carroll's prior $%
\widetilde{\pi }$ is a monotonic strengthening of Bob's prior $\widehat{\pi }
$ (and we make no assumptions relating to Alice). Then Proposition \ref%
{p:upper} (iii) implies that\footnote{%
Apply Proposition \ref{p:upper} with $\widehat{\pi }$ instead of $\pi $ and
use (iii) with $R=Q_{\Gamma }$ (which is Alice's posterior probability of $%
\Gamma $, and is increasing in $s).$ We need to use (iii) because monotonic
optimism here means $(\widehat{Q}_{\Gamma },\widetilde{\mathbb{P}})\geq _{%
\mathrm{lr}}(\widehat{Q}_{\Gamma },\widehat{\mathbb{P}})$, where $\widehat{Q}%
_{\Gamma }$ denotes Bob's posterior probability of $\Gamma $.} $(Q_{\Gamma },%
\widetilde{\mathbb{P}})\geq _{\mathrm{lr}}(Q_{\Gamma },\widehat{\mathbb{P}})$
for any upper set $\Gamma $ and any MLRP signaling structure: Carroll is
more optimistic than Bob about Alice's posterior probabilities---and thus,
about the market price.